\documentclass{iopart}

\usepackage{ifthen}
\usepackage{ifpdf}

\usepackage{latexsym}
\usepackage{amssymb}
\usepackage{bm}

\ifpdf
\usepackage{graphicx}
\usepackage{epstopdf}
\else
\usepackage{graphicx}
\usepackage{epsfig}
\fi

\newcommand{\const}{\mbox{const}}
\newcommand{\trc}{\mbox{trace}}

\newcommand{\im}{\mbox{Im}}

\newcommand{\eexp}{\mbox{e}^}

\newcommand{\tbox}[1]{\mbox{\tiny #1}}

\newcommand{\amatrix}[1]{\matrix{#1}}
\newcommand{\beq}[1]{\begin{eqnarray}\ifthenelse{#1=-1}{\nonumber}
{\ifthenelse{#1=0}{}{\label{e#1}}}}
\newcommand{\eeq}{\end{eqnarray}}
\newcommand{\hide}[1]{}

\newcommand{\putgraph}[2][width=0.30\hsize]{\includegraphics[#1]{#2}}

\begin{document}

\title{Dephasing of a particle in a dissipative environment}
%\shorttitle{Dephasing}

\author{Doron Cohen and Baruch Horovitz}

\address{Department of Physics, Ben-Gurion University, Beer-Sheva 84105, Israel}

\begin{abstract}
The motion of a particle in a ring of length $L$
is influenced by a dirty metal environment whose
fluctuations are characterized by a short correlation
distance ${\ell \ll L}$.
We analyze the induced decoherence process, and compare
the results with those obtained in the opposing
Caldeira-Leggett limit (${\ell \gg L}$).
A proper definition of the dephasing factor that does
not depend on a vague semiclassical picture is employed.
Some recent Monte-Carlo results about the effect
of finite temperatures on ``mass renormalization"
in this system are illuminated.
\end{abstract}

%%%%%%%%%%%%%%%%%%%%%%%%%%%%%%%%%%%%%%%%%%%%%%%%%%%%%%%%%%%%%%
%%%%%%%%%%%%%%%%%%%%%%%%%%%%%%%%%%%%%%%%%%%%%%%%%%%%%%%%%%%%%%
%%%%%%%%%%%%%%%%%%%%%%%%%%%%%%%%%%%%%%%%%%%%%%%%%%%%%%%%%%%%%%
%%%%%%%%%%%%%%%%%%%%%%%%%%%%%%%%%%%%%%%%%%%%%%%%%%%%%%%%%%%%%%

%%%%%%%%%%%%%%%%%%%%%%%%%%%%%%%%%%%%%%%%%%%%%%%%%%%%%%%%%%%%%%%%
\section{Introduction}

What is the dephasing of a particle that has an energy~$E$ if it is subject
to a fluctuating  environment that has a temperature~$T$?
In particular what is the dephasing close to equilibrium ($E\sim T$),
and what happens in the limit $T\rightarrow0$?
This question has fascinated the mesoscopic community during the last
two decades \cite{AAK,imry,webb,zaikin,alt,golub,delft}.
Our purpose is to study this question within the framework of
linear response theory for a general characterization of the environment.
In the Caldeira-Leggett (CL) framework \cite{FV,CL}\footnote{We emphasize  
here the {\em Ohmic} CL model, which is of relevance  
in the present context as a limiting case for a dirty metal environment. 
Obviously in general one may consider non-Ohmic models, 
where memory kernels are involved while the notion of 
a friction constant~$\eta$ becomes ill defined.}     
the effect of the environment is characterized by a
friction coefficient~$\eta$ and by a temperature~$T$. But more
generally \cite{dld,qbm,dph} it has been emphasized that the
proper way to characterize the environment is by its form factor
$\tilde{S}(q,\omega)$. The form factor contains information on
both the temporal and the spatial aspects of the fluctuations, and
in particular one can extract from it not only $T$ and $\eta$, but
also the spatial correlations. Typically (but not always) these
spatial correlations can be characterized
by a correlation distance $\ell$.

So now we ask the refined question: Given $\tilde{S}(q,\omega)$,
what is the dephasing? But first we have to say what do we mean by
dephasing. In Ref.\cite{dld,qbm} the CL  approach
has been generalized. Namely, it has been realized that an
environment with a given $\tilde{S}(q,\omega)$ can be modeled as a
set of Harmonic oscillators. Then it is possible to apply the
Feynman-Vernon formalism in order to trace them out. Using a
semiclassical point of view the propagator is expressed
as a sum over pairs of classical trajectories.
One observes that after time~$t$ the interference contribution
(from the off diagonal terms in the double sum) is suppressed
by a factor $P_{\varphi}$. This factor is interpreted as
a dephasing factor, and its expression can be cast into the
form ${P_{\varphi}=\exp(-F(t))}$ with
\beq{1}
F(t) \ \ = \ \  \int d\bm{q} \int \frac{d\omega}{2\pi} \,
\tilde{S}(\bm{q},\omega) \, \tilde{P}(-\bm{q},-\omega; t) \,t \,\,
\eeq
where the $d\bm{q}$ integration measure depends of the dimensionality.
In the semiclassical treatment $\tilde{S}(q,\omega)$
would be the {\em symmetrized} form factor of the environment and
$\tilde{P}(q,\omega;t)$ would be the classical {\em symmetric} power
spectrum of the motion. There are some subtleties in the
definition of $\tilde{P}(q,\omega;t)$ that we are going to discuss
later on. In particular we note that $\tilde{P}(q,\omega;t)$ may
have weak dependence on~$t$ because a finite time interval is considered.

It has been further argued in Ref.\cite{dph} that due to inherent
limitations of the semiclassical (stationary phase) approximation
the physically correct procedure is to use the  {\em non}-symmetrized
quantum versions of $\tilde{S}(q,\omega)$ and $\tilde{P}(q,\omega;t)$.
This point has been further discussed in \cite{florian}.
However, a proper derivation of Eq.(\ref{e1}), that does
not rely on the semiclassical framework, has not been introduced.
One objective of the present paper is to extend the derivation
of the above formula beyond the semiclassical context.

It is important to realize that Eq.(\ref{e1}) is capable of
reproducing all the established results about dephasing in
normal metals, including the high temperature~$\propto T$
dependence of the dephasing time, the low temperature $\propto
T^{3/2}$ dependence in the case of a diffusive particle, and the
$\propto T^2$ dependence in the ballistic regime. At finite
temperatures there is a finite time scale $\hbar/T$ that allows
the the approximation $F(t)\approx\Gamma_{\varphi} t$ and hence
the notion of dephasing rate $\Gamma_{\varphi}$ is well defined.
In the limit $T\rightarrow0$ we always have
$\Gamma_{\varphi}\rightarrow0$. This, however, does not
exclude sub-exponential (power law) decay of $P_{\varphi}$.

Indeed it is well known \cite{golub} that for a
Brownian particle with CL environment the function $F(t)$
grows as $\log(t)$ at zero temperature implying
sub-exponential dephasing at $T=0$. It is a common miss-conception
that Eq.(\ref{e1}) with {\em non}-symmetrized spectral functions
fails to reproduce this $\log(t)$ spreading. We shall dwell
on this point later on in this paper. Furthermore we shall
study whether similar sub-diffusive behavior can be found
for general $\tilde{S}(q,\omega)$.

During the last decade the study of a particle in a ring, has
become a paradigm for the study of ground state anomalies.
\cite{Buttiker1,Buttiker2,guinea,golubev,horovitz}. Besides being
a prototype problem that possibly can be realized as a
mesoscopic electronic device, it is also of relevance to
experiments with particles that are trapped above an ``atom chip"
device \cite{harber,jones,lin}, where noise is induced by nearby
metal surfaces.
A significant progress has been achieved in analyzing the
equilibrium properties of this prototype system, in particular the
dependence of the ground state energy on the Aharonov Bohm flux
through the ring. The derivations of the dephasing
factor using Eq.(\ref{e1}) for the ring problem is a major
objective of the present paper.
In this context there is a growing understanding that
the study of dephasing is intimately connected with the study
of mass-renormalization at low temperatures. We believe that
our results shed new light on some recent findings~\cite{horovitz2}
that have been obtained using Monte-Carlo data for the temperature dependence
of the mass-renormalization effect.

%%%%%%%%%%%%%%%%%%%%%%%%%%%%%%%%%%%%%%%%%%%%%%%%%%%%%%%%%%%%%%%%

\ \\

{\bf Outline:} In sections~2-3 we characterize the environment by
the power spectrum of its fluctuations, and then in Sections~4-6
we derive the formula for the dephasing factor. This formula is
applied in Sections~7-8 to the calculation of the dephasing of a
particle of mass $M$ in a ring of length $L$. The results depends
crucially on the correlation distance $\ell$ of the
fluctuating environment. They shed light on some new findings
regarding mass-renormalization in this system as explained in
Section~9. It is conjectured that the mass-renormalization effect
involves a measure for coherence. Some further discussion
of the theoretical framework is presented in Section~10.

%%%%%%%%%%%%%%%%%%%%%%%%%%%%%%%%%%%%%%%%%%%%%%%%%%%%%%%%%%%%%%%%
\section{The characterization of a fluctuating field}

We regard the environment as a fluctuating field~$\mathcal{U}(x,t)$.
See appendix~A for more details on its Hamiltonian modeling.
The fluctuations of the environment
are characterize by a form factor:
\beq{3}
\tilde{S}(q,\omega) =
\int \int
\Big[ \big\langle \hat{\mathcal{U}}(x,t)
\hat{\mathcal{U}}(0,0) \big\rangle \Big]
\, \eexp{i\omega t-iqx} \, dtdx
\eeq
where the expectation value assumes that the bath is in a stationary
state of its unperturbed Hamiltonian.
%%%
%%% and $\mbox{FT}_{q,\omega}$ is the
%%% Fourier transform with respect
%%% to the separation ${r=x_2-x_1}$
%%% and the time difference ${\tau=t_2-t_1}$]].
%%%
The force operator is formally defined as ${\mathcal{F}=-\mathcal{U}'(x,t)}$,
where the derivative is taken with respect to~$x$. The force-force
correlation function is obtained via double differentiation of the
correlation function. In particular the local power spectrum of
the fluctuating force is
\beq{4}
\tilde{S}(\omega) = \int \frac{dq}{2\pi} q^2 \tilde{S}(q,\omega)
\eeq
An Ohmic environment is characterized by
\beq{5}
\tilde{S}_{\tbox{ohmic}}(\omega)=\frac{2\hbar\eta\omega}{1-\eexp{-\hbar\omega/T}}
\ \ \ \ \ \ \mbox{for}
\ \ |\omega|<\omega_c
\eeq
The friction coefficient characterizes the response
of the environment to an $x$ variation (``force proportional to velocity").
Accordingly it is given by the Kubo formula:
\beq{6}
\eta \ \ = \ \ \lim_{\omega\rightarrow 0}
\frac{1}{2\hbar\omega}
\Big[\tilde{S}(\omega)-\tilde{S}(-\omega)\Big] \ \ = \ \
\frac{\tilde{S}(\omega{=}0)}{2T}
\eeq
For a strictly Ohmic bath the friction coefficient is frequency
independent and the first equality holds for any
${\omega<\omega_c}$ (no need to take a limit). 
The generalization of the above to 3~dimensions is straight forward.
The position coordinate becomes $\bm{x}=(x,y,z)$ and accordingly
$q$ should be replaced by $\bm{q}=(q_x,q_y,q_z)$, with integration
measure $d^3\bm{q}/(2\pi)^3$. In the definition of
$\tilde{S}(\omega)$ the $q^2$ should be replaced by $q_x^2$ or
optionally by $\bm{q}^2/3$. The simplest type of environment is
known as the CL model, where the particle interacts with long
wavelength modes. The associated form factor is
\beq{7}
\tilde{S}_{\tbox{CaldeiraLeggett}}(q,\omega)
\ \ = \ \ \tilde{S}_{\tbox{ohmic}}(\omega) \times 3 \frac{(2\pi)^3 \delta^3(\bm{q})}{\bm{q}^2}
\eeq
Another case of interest is the interaction with a dirty metal (Appendix~B) for which
\beq{8}
\tilde{S}_{\tbox{DirtyMetal}}(q,\omega)
\ \ \approx \ \
\tilde{S}_{\tbox{ohmic}}(\omega)
\times \frac{4\pi\ell^3}{\bm{q}^2}
\ \ \ \ \ \ \mbox{for}
\ \  |\bm{q}|\lesssim\frac{1}{\ell}
\eeq
where the friction coefficient
can be expressed in terms of the conductivity:
\beq{9}
\eta \ \ = \ \ \frac{e^2}{\sigma}\times \frac{1}{4\pi\ell^3}
\eeq
In the latter context it is customary to define a
dimensionless friction coefficient as follows:
\beq{10}
\alpha \ \ = \ \ \frac{1}{2\pi}\eta\ell^2
\ \ = \ \ \frac{e^2}{8\pi^2\sigma\ell}
\ \ = \ \ \frac{3}{8(k_F\ell)^{2}}
\eeq
The motion of a classical Brownian particle of mass $M$ under
the influence of such fluctuating environment is characterized by a damping rate
\beq{1050}
\gamma \ \ = \ \ \frac{\eta}{M} \ \ = \ \ \frac{2\pi\alpha}{M\ell^2}
\eeq

%%%%%%%%%%%%%%%%%%%%%%%%%%%%%%%%%%%%%%%%%%%%%%%%%%%%%%%%%%%%%%%%
\section{The fluctuations within a ring}

In the present paper we consider a particle in a ring of radius
$L/(2\pi)$. We assume that
$\tilde{S}(\bm{q},\omega)=\tilde{S}_{\tbox{ohmic}}(\omega)\tilde{w}(\bm{q})$
is factorisable, as in the examples of Eqs. (\ref{e7},\ref{e8}).
Since the motion is confined to one dimension it is natural to
expand the spatial correlations along the rings in Fourier series:
\beq{10001}
\int \frac{d^3q}{(2\pi)^3}
\,\tilde{w}(\bm{q})
\,\eexp{i\bm{q}\cdot [\bm{R}(\theta_2)-\bm{R}(\theta_1)]}
\ \ = \ \
\sum_{m=-\infty}^{\infty} w_m\eexp{im(\theta_2-\theta_1)}
\eeq
Accordingly, using Eq.(\ref{e3}) with $x=(L/2\pi)\theta$ we get
\beq{11}
\tilde{S}(q,\omega)
\ \ = \ \
\tilde{S}_{\tbox{ohmic}}(\omega)
\times \sum_{m=-\infty}^{\infty} w_m \, 2\pi\delta(q-q_m)
\eeq
where the discrete modes are
\beq{12}
q_m \ \ = \ \ \frac{2\pi}{L}m,
\ \ \ \ \ \ \ \ \ \ \ m=0,\pm1,\pm2,...
\eeq
By convention we want $\eta$ to be the friction coefficient.
Therefore $\tilde{S}(\omega)$ as defined by Eq.(\ref{e4}) should
equal~$\tilde{S}_{\tbox{ohmic}}(\omega)$ of Eq.(\ref{e5}).
This implies the following sum rule:
\beq{13}
\sum_{m=-\infty}^{\infty} w_m q_m^2 = 1
\eeq
In general we have $\sim (L/\ell)$  fluctuating modes,
each has the weight $w_n \sim \ell^3/L$.
In Appendix~C we show that for a CL bath
we have only one fluctuating mode ($|m|=1$) with
\beq{14}
w_m = \frac{1}{2} \left(\frac{L}{2\pi}\right)^2
\eeq
while in the case of a Dirty metal with short range
correlated fluctuations we have ${\mathcal{M}=(L/(2\pi))/\ell \gg 1}$
fluctuating modes with weights
\beq{15}
w_m \approx
\frac{\ell^2}{2\pi} \times \frac{1}{\mathcal{M}} \,\ln\left(\frac{\mathcal{M}}{|m|}\right)
\ \ \ \ \ \ \ \ \ \ \ \mbox{for $|m|<\mathcal{M}$}
\eeq
In both cases we ignore the $m{=}0$ mode for a reason which
is explained in the next section. It is important to realize that
the CL model can formally be regarded as a special limit of a
dirty metal environment with ${\ell\gg L}$. In the latter case the
weight of the ${|m|>1}$ modes is smaller by powers
of $L/\ell$ (Appendix C).

%%%%%%%%%%%%%%%%%%%%%%%%%%%%%%%%%%%%%%%%%%%%%%%%%%%%%%%%%%%%%%%%
\section{The dephasing factor}

The dephasing factor $P_{\varphi}$ is a number within $[0,1]$ that
characterizes the suppression of coherence. Its popular definition
is based on a semiclassical picture.
Using the Feynman-Vernon formalism the propagator is expressed
as a sum over pairs of classical trajectories.
One observes that after time~$t$ the interference contribution
(from the off diagonal terms in the double sum) is suppressed
by a factor
\beq{16}
P_{\varphi}(t)
\ \ = \ \ \Big| \ \Big\langle \ U[x^A] \chi \ \Big| \ U[x^B] \chi \ \Big\rangle \ \Big|
\ \ = \ \ \eexp{-S_N[x^A,x^B]}
\eeq
where $\chi$ is the preparation of the 
bath\footnote{The implicit assumption of initial 
factorized state is of course problematic \cite{sols,weiss}. 
In most cases it implies an unpleasant transient 
that should be ignored. We further discuss 
the significance of the long time decay later 
in this section after Eq.(\ref{e17}) and in the Summary}.  
In order not to complicate the notations, the canonical average
over $\chi$ states is implicit.
The unitary operator $U[x]$ generates the evolution of the bath given
that the particle goes along the trajectory~$x(t)$. The action
$S_N[x^A,x^B]$ is a double time integral. Using manipulation as in
Ref.\cite{qbm,dph} one obtains Eq.(\ref{e1}) with the symmetrized
version of $\tilde{S}(q,\omega)$, and the symmetric classical
version of $\tilde{P}(q,\omega)$. This semiclassical expression is
definitely wrong for short range scattering at low temperatures
\cite{dph}, because it does not reflect that {\em closed channels
cannot be excited}. This problem with the semiclassical
(stationary phase) approximation is well known in the theory of
inelastic scattering.
One way to overcome this limitation is to refine the definition
of the dephasing factor using a semiclassically inspired
``scattering" point of view as in Appendix~D. However it is clear
that such a refinement makes the concept of dephasing too vague.

We propose here a natural definition for the dephasing factor
that is related to the purity $\trc(\rho^2)$ of the reduced
probability matrix. The notion of purity is very old, but in
recent years it has become very popular due to the interest in
quantum computation \cite{purity}. Given that the state of the
system including the environment is $\Psi_{pn}$, where $p$ and $n$
label the basis states of the particle and the bath respectively, 
the purity is given by
\beq{17}
P_{\varphi}(t)
\ \ &=& \ \ \sqrt{\trc(\rho_{\tbox{sys}}^2)}
\ \ = \ \ \sqrt{\trc(\rho_{\tbox{env}}^2)}
\nonumber\\
\ \ &=& \ \ \left[\sum_{p'p''n'n''} \Psi_{p'n'}\Psi_{p''n'}^*\Psi_{p''n''}\Psi_{p'n''}^*\right]^{1/2}
\eeq
Assuming a factorized initial preparation as in the conventional
Feynman-Vernon formalism, we propose the loss of purity (${P_{\varphi}<1}$) 
as a measure for decoherence. A standard reservation applies:  
initial transients during which the system gets ``dressed" by the environment 
should be ignored as these reflect renormalizations
due to the interactions with the high frequency modes.
Other choices of initial state might involve
different transients, while the later slow approach
to equilibrium should be independent of these transients.
In any case the reasoning here is not much different
from the usual ideology of the Fermi golden rule,
which is used with similar restrictions to calculate
transition rates between levels.

Writing the initial preparation as ${\Psi^{(0)}_{pn}=\delta_{p,p_0}\delta_{n,n_0}}$,
and using leading order perturbation theory, we can relate $P_{\varphi}$ to the
probabilities ${P_t(p,n|p_0,n_0)=|\Psi_{pn}|^2}$ to have a transition
from the state $|p_0,n_0\rangle$ to the state $|p,n\rangle$ after time~$t$.
The derivation is detailed in Appendix~E. One obtains the result
\beq{18}
\hspace*{-15mm}
P_{\varphi}(t) = P_t(p_0,n_0|p_0,n_0) + P_t(p{\ne}p_0,n_0|p_0,n_0) +  P_t(p_0,n{\ne}n_0|p_0,n_0)
\eeq
in agreement with the semiclassically inspired point of view of
Appendix~D. The notation $p\neq p_0$ or $n\neq n_0$ implies a
summation $\sum_{p\neq p_0}$ or $\sum_{n\neq n_0}$, respectively.
In the next section we shall discuss the actual calculation
of $P_t(p,n|p_0,n_0)$, using the Fermi-golden-rule (FGR). Thus we
deduce that within the FGR framework, the purity is simply the probability
that either the system or the bath do not make a transition.
The first term in Eq.(\ref{e18}) is just the survival probability
of the preparation $P_{\tbox{survival}}=P_t(p_0,n_0|p_0,n_0)$.
The importance of the two other terms can be demonstrated using simple
examples: For an environment that consists of static scatterers
we have ${P_{\tbox{survival}}<1}$ but ${P_{\varphi} = 1}$
thanks to the second term. For a particle in a ring that interacts
with a $q{=}0$ environmental mode ${P_{\tbox{survival}}<1}$
but ${P_{\varphi} = 1}$ thanks to the third term.
Using $\sum_{p,n}P_t(p,n|p_0,n_0)=1$ we obtain the optional expression
\beq{10002}
p_{\varphi} \ \ = \ \ 1-P_{\varphi} \ \ = \ \
\sum_{p\neq p_0}\sum_{n\neq n_0}P_t(p,n|p_0,n_0)
\eeq
In the problem that we consider in this paper we can calculate $P_{\varphi}$
using a ${dq d\omega}$ integral as in Eq.(\ref{e1}). In many examples
the ${\omega=0}$ transitions have zero measure and therefore $P_{\varphi}$ is
practically the same as $P_t(p_0,n_0|p_0,n_0)$.
Otherwise one has to be careful in eliminating those transitions that do
not contribute to the dephasing process.
Anticipating the application of Eq.(\ref{e1}) for the calculation
of the dephasing for a particle in a ring, the integration over~$q$ becomes
a summation over~$q_m$, and the $m=0$ component should be excluded.

%%%%%%%%%%%%%%%%%%%%%%%%%%%%%%%%%%%%%%%%%%%%%%%%%%%%%%%%%%%%%%%%
\section{Dephasing at finite temperatures}

The interaction between the particle ($\hat{x}$)
and the environment can be written as in Appendix~A:
\beq{19}
V \ \ = \ \  \int dx \, \hat{\rho}(x) \, \hat{\mathcal{U}}(x)
\eeq
where $\hat{\rho}(x)=\delta(x-\hat{x})$. In the Heisenberg (interaction) picture
a time index is added so we have $\hat{\mathcal{U}}(x,t)$ and $\hat{\rho}(x,t)$.
Given a preparation of the bath and of the system we can define $\tilde{S}(q,\omega)$
to characterize $\hat{\mathcal{U}}(x,t)$ and we can also
define $\tilde{P}(q,\omega)$ to characterize $\hat{\rho}(x,t)$.
The precise definition of the latter object is further discussed below.
The survival probability of a factorized preparation
is ${P_{\varphi}(t)=1{-}p_{\varphi}(t)}$ where:
\beq{20}
\hspace{-2cm}
p_{\varphi}(t)
\ \ = \ \
\int_0^t \int_0^t
\langle V(t_2) V(t_1) \rangle
\, dt_2dt_1
\\ \nonumber
\hspace{-2cm}
\ \ = \ \
\int\!\!\!\!\!\int dt_1dt_2 \int\!\!\!\!\!\int dx_1 dx_2 \,
\Big\langle  \rho(x_2,t_2)\mathcal{U}(x_2,t_2) \, \rho(x_1,t_1)\mathcal{U}(x_1,t_1) \Big\rangle
\\ \nonumber
\hspace{-2cm}
\ \ = \ \
\int\!\!\!\!\!\int \frac{dq}{2\pi}\frac{d\omega}{2\pi} \tilde{S}(q,\omega)
\int\!\!\!\!\!\int dt_1dt_2 \int\!\!\!\!\!\int dx_1 dx_2 \,
\langle \rho(x_2,t_2) \rho(x_1,t_1) \rangle
\, \eexp{iq(x_2-x_1)-i\omega(t_2-t_1)}
\eeq
At finite temperatures, if recurrences due to finite-size
quantization effect can be ignored, one can obtain as an
approximation ${p_{\varphi}\approx\Gamma_{\varphi} t}$,
where $\Gamma_{\varphi}$ is called the dephasing rate.
In the next section we discuss circumstances
where such an approximation is not valid:
the feasibility of this approximation requires neglect
of the end-point contributions to the double time
integration. By going to the variables $(t_1+t_2)/2$
and ${\tau=t_2-t_1}$ one obtains the following expression
for the dephasing rate:
\beq{21}
\Gamma_{\varphi} \ \ = \ \ \int\!\!\!\!\!\int \frac{dq}{2\pi}\frac{d\omega}{2\pi}
\, \tilde{S}(q,\omega) \, \tilde{P}(-q,-\omega)
\eeq
The implied definition of $\tilde{P}(q,\omega)$ is
discussed below and further refined in the next section.
Following standard argumentation one conjectures that
the long time decay of $P_{\varphi}(t)$ is exponential,
as in the analysis of Wigner's decay.
The similarity of Eq.(\ref{e21}) to the semiclassical
result (as discussed below Eq.(\ref{e1})) is obvious.
It is important to realize that
in the present context the {\em non}-symmetrized quantum
version of the power spectrum has emerged.
Furthermore, if we want to calculate~$P_{\varphi}$,
and not just the survival probability of the initial state,
we have to be careful about the proper treatment
of the diagonal terms as discussed in the previous section.
Accordingly we eliminate the diagonal term from the
implied definition of the power spectrum:
\beq{22}
\tilde{P}(q,\omega) =
\int_{-\infty}^{+\infty}
\Big[
\langle \eexp{-iq x(\tau)} \eexp{iq x(0)} \rangle
- \langle \eexp{iq x}\rangle^2
\Big]
\, \eexp{i\omega \tau} \, d\tau
\eeq
We emphasize again that in a later section we 
are going to treat the time limits 
more carefully, where ${\tilde{P}(q,\omega)}$ 
will be replaced by ${\tilde{P}(q,\omega;t)}$ 
as in Eq.(\ref{e1}). 
For a ballistic particle with mass $M$ and
momentum $p = (2M E)^{1/2}$ we have:
\beq{23}
\tilde{P}(q,\omega) \ \ = \ \ 2\pi\delta(\omega-\omega(q))
\eeq
where $\omega(q)=[(p+q)^2-p^2]/(2M)$. The power spectrum is
illustrated in Fig.~1. The expectation value in Eq.(\ref{e23}) 
is taken for a particle with momentum ${p}$.  
For the ground state ${p{=}0}$ and hence $\omega(q)=q^2/2M$. 
In particular for a particle on a ring Eq.(\ref{e11}) 
implies $\omega(q_m)=q_m^2/2M$. The ballistic case should be
contrasted (see Fig.~1) with the power spectrum of a diffusive
particle:
\beq{24}
\tilde{P}(q,\omega) \ \ = \ \ \frac{2Dq^2}{\omega^2+(Dq^2)^2}
\eeq
where $D$ is the diffusion coefficient. In the ballistic case the
power spectrum is concentrated along $\omega=\omega(q)$, while in
the diffusive case it spreads over the range ${|\omega| < Dq^2}$.
In any case
\beq{25}
\int_{\omega{\ne}0} \tilde{P}(q,\omega) \frac{d\omega}{2\pi} \ \ = \ \ 1
\ \ \ \ \ \ \ \ \ \ \ \ \mbox{by definition, for any $q$}
\eeq
Assuming close-to-equilibrium conditions, the expectation
value in Eq.(\ref{e22}) should reflect a thermal state
with energy ${E \sim T}$. In practice one may set ${E \sim 0}$,
though in general better to be careful about it:
looking at Fig.~1 one can deduce that to take ${E \sim 0}$,
in the problem that we are going to consider,
results in an underestimation of the dephasing rate
by a~$\sqrt{2}$ factor\footnote{Assuming that ${E \sim 0}$ we are going 
to explain in Section~7 that the non-negligible 
contribution to the integral comes form the range $|q|<q_T$ 
of effective modes: only the fluctuating modes 
in the rectangular region of Fig.~1 
resonate with the particle and hence contribute. 
The power spectrum of the particle for ${E \sim T}$ 
is shifted ``upwards" in~$\omega$ and consequently 
the effective $q$~range becomes larger by factor $\sqrt{2}$ 
compared with the ${E \sim 0}$ case.}.

%%%%%%%%%%%%%%%%%%%%%%%%%%%%%%%%%%%%%%%%%%%%%%%%%%%%%%%%%%%%%%%%
%%%%%%%%%%%%%%%%%%%%%%%%%%%%%%%%%%%%%%%%%%%%%%%%%%%%%%%%%%%%%%%%
\section{Dephasing at ``zero" temperature}

The expression for $\Gamma_{\varphi}$ gives manifestly zero
dephasing rate in the limit of zero temperature, because in this
limit $\tilde{S}(q,\omega)$ and $\tilde{P}(-q,-\omega)$ have no
overlap. However, this does not mean that $P_{\varphi}$ does not
decay. It still might have a sub-exponential decay. In order to
understand this point we first discuss a simple artificial
calculation of the double time integral
\beq{30}
p_{\varphi}(t) = \int_0^t\int_0^t C(t_1-t_2) \, dt_1 dt_2
\eeq
where $C(\tau)$ is the symmetrized force-force correlation:
its Fourier transform $\tilde{C}(\omega)$ is the
symmetrized version of $\tilde{S}(\omega)$ of Eq.(\ref{e5}).
Later in this section we come back to the actual calculation
and discuss how Eq.(\ref{e20}) can be treated.

The approximation $p_{\varphi}(t)\approx\Gamma_{\varphi} t$
for the integral in Eq.(\ref{e30}) is based on the assumption
that $C(\tau)$ has short range correlations with non-vanishing integral.
At finite $T$ this assumption is indeed satisfied
because ${\tilde{S}(\omega{=}0)=2\eta T}$ is finite.
But at zero temperature the integral over $C(\tau)$ is zero.
In fact at zero temperature Eq.(\ref{e5}) implies
that $\tilde{C}(\omega) = 2\eta|\omega|$ and
hence $C(\tau)$ has power law tails $-(\eta/\pi)/\tau^2$.
It is important to realize that the $T{=}0$
behavior prevails also at finite temperatures
provided ${T<\omega_c}$, and the time of interest
should be smaller compared with $1/T$.
Under such ``$T{=}0$" conditions $C(\tau)$ can
be approximated by its $T{=}0$ version.
In order to see what comes out from Eq.(\ref{e30}) we observe that
\beq{31}
\Gamma_{\varphi}(t) = \frac{d}{dt}p_{\varphi}(t) = \int_{-t}^{t} C(\tau) \, d\tau
\eeq
For an Ohmic bath at ``zero temperature" the integral
over the power law tails of $C(\tau)$ gives $\Gamma(t) \propto 1/t$,
hence the spreading is logarithmic:
\beq{0}
p_{\varphi}(t) = \frac{2}{\pi}\eta\,\ln(\omega_c t) + \const
\eeq
It is instructive to make the same
calculation in $\omega$ space. One realizes that
\beq{0}
p_{\varphi}(t) = \int \frac{d\omega}{2\pi} \tilde{C}(\omega) \left[\frac{\sin(\omega t /2)}{\omega/2} \right]^2
\eeq
which for $\tilde{C}(\omega) \propto |\omega|$
gives correctly the logarithmic spreading.

Without any approximation we can generalize the above
treatment so as to handle Eq.(\ref{e20}),
taking also into account the {\em non}-symmetrized nature
of the spectral functions. 
Performing the $dx_1dx_2$ integration  
we obtain ${p_{\varphi}(t) = F(t)}$ 
as in Eq.(\ref{e1}) where
%
% \beq{1001}
\beq{0}
\tilde{P}(q,\omega; t) \ \ = \ \ 
\frac{1}{t}
\int_{0}^{t}
\int_{0}^{t}
\langle \eexp{-iq x(t_2)} \eexp{iq x(t_1)} \rangle
\, \eexp{i\omega (t_2-t_1)} \, dt_1 dt_2
\eeq
In complete analogy with the way in which the 
environmental fluctuations has been treated, 
we express the correlator as a Fourier integral  
over ${\tilde{P}(q,\omega)}$, and then we are able 
to explicitly perform the $dt_1dt_2$ integration.
The outcome of this procedure allows to express 
the result as a convolution:
\beq{33} 
\tilde{P}(q,\omega;t) \ \ = \ \ 
\frac{1}{2\pi t}
\left[\frac{\sin(\omega t /2)}{\omega/2} \right]^2
(*) \,\tilde{P}(q,\omega) 
\eeq
An optional compact way of writing the final result is 
\beq{1001}
\hspace*{-15mm} p_{\varphi}(t) =  \int d\bm{q}
\int\!\!\!\!\! \int\frac{d\omega}{2\pi}\frac{d\omega'}{2\pi}
\tilde{S}(\bm{q},\omega)   \tilde{P}(-\bm{q},-\omega')
\left[\frac{\sin((\omega{-}\omega')t /2)}{(\omega{-}\omega')/2} \right]^2
\eeq
We note that in practical calculations  
or for aesthetic reasons it is possible 
to make the replacement
\beq{32} 
\left[\frac{\sin(\omega t /2)}{\omega/2} \right]^2 \ \
\longmapsto \ \ \left[\frac{(2/t)}{(1/t)^2+\omega^2}\right] \times t 
\ \ \ \ \ \ \ \ \ \ \ \ \mbox{[optional]}
\eeq
The more convenient Lorentzian kernel has the same normalization, 
the same width~$1/t$, and the same ${2/\omega^2}$ tails. 
It can be regarded as arising from using ``soft" 
rather than ``sharp" cutoff for the time integration.

%%%%%%%%%%%%%%%%%%%%%%%%%%%%%%%%%%%%%%%%%%%%%%%%%%%%%%%%%%%%%%%%
%%%%%%%%%%%%%%%%%%%%%%%%%%%%%%%%%%%%%%%%%%%%%%%%%%%%%%%%%%%%%%%%
\section{Dephasing in the presence of a dirty metal ``$T{>}0$"}

We turn to consider a particle of mass~$M$ in a ring of length~$L$.
We assume close-to-equilibrium conditions so we take
the energy of the particle above the ground state as ${E\sim T}$.
We consider in this section temperatures~$T$ that are
much larger compared with the level spacing~${\Delta \sim (ML^2)^{-1}}$.
This does not mean that the system is not 
coherent\footnote{As discussed in section~9 the coherence 
measure is ${\Gamma_{\varphi}/\Delta_{\tbox{eff}}}$. 
In the regime of main interest ${\Gamma_{\varphi} \ll T}$. 
Furthermore, the relevant energy scale is not necessarily 
the level spacing: in the mesoscopic context the
relevant energy scale (e.g. the ``Thouless energy") 
is typically much larger and proportional to $\hbar$ 
in contrast to the microscopic quantization scale which 
is proportional to $\hbar^{\tbox{dimensionality}}$.}.    
We further assume that the time of interest is much longer
compared with the relevant dynamical time scales,
and in particular compared with~${1/T}$.
With this assumption it is legitimate to use Eq.(\ref{e21})
to calculate the dephasing rate $\Gamma_{\varphi}$,
and to  treat the $d\omega$ integration
as if the levels of the ring form a continuum.
With the substitution of Eqs.(\ref{e11},\ref{e23}) this
leads to the following result:
\beq{1004}
\Gamma_{\varphi} \ \ = \ \ \sum_m w_m \,\tilde{S}_{\tbox{ohmic}}(-\omega(q_m))
\eeq
Of course one has to verify at the end of the calculation the
self-consistency condition ${\Gamma_{\varphi} \ll \min\{T,\omega_c\}}$.
This condition would be satisfied if the system-environment coupling
is not too strong.

A graphic illustration of the ${(q,\omega)}$ integration domain
is presented in Fig.~1. The $T$ dependence of $\tilde{S}_{\tbox{ohmic}}$
in Eq.(\ref{e1004}) limits the sum to $\omega(q_m)<T$.
Taking into account the weight factors the effective number
of fluctuating modes is
\beq{26001}
\mathcal{M}_{\tbox{eff}} \ \ = \ \
\frac{L}{2\pi} q_{\tbox{eff}}
\ \ \approx \ \
\min\left\{\,\, \mathcal{M}, \,\, q_cL, \,\,  q_TL   \,\, \right\}
\eeq
where $q_c = (2M \omega_c)^{1/2}$ and $q_T = (2M T)^{1/2}$. Note
that if we had Fermi occupation it would be $q_T = T/v_F$, while
for diffusive motion it would be $q_T = (T/D)^{1/2}$.
The dephasing rate is obtained by summing over all
the contributing modes. Each effective mode
contributes $2\eta T \times w_m$ to the sum. Accordingly
\beq{26}
\Gamma_{\varphi} \ \ = \ \ 2\eta T \times \sum_{0<|q_m|<q_{\tbox{eff}}} w_m
\ \ \sim \ \  2\eta T \times \bar{w} \mathcal{M}_{\tbox{eff}}
\eeq
where the average weight
is $\bar{w} \sim \ell^2/\mathcal{M}\sim \ell^3/L$
for a short range correlated dirty metal
environment (${\ell \ll L}$), while $\bar{w} \sim L^2$ in the
opposite CL limit (${\ell \gg L}$), as implied by
Eq.(\ref{e15}) and Eq.(\ref{e14}) respectively.
If all the modes are effective
we get ${\Gamma_{\varphi}=2\eta T \times \ell^2}$,
while in the case of a CL environment
we get the well known result
\beq{27}
\Gamma_{\varphi} \ \ = \ \ 2\eta T \times L^2
\ \ \ \ \ \ \ \ \Big[\mbox{Caldeira-Leggett}\Big]
\eeq
For a fluctuating environment with correlation distance $\ell$,
Eq.(\ref{e26001}) implies a crossover temperature:
\beq{1051}
T^{*} \ \ = \ \ \min\left\{\, \frac{1}{M\ell^2},\ \omega_c \, \right\}
\eeq
For $T<T^{*}$ the dephasing rate depends on $q_T$ and therefore
develops non-linear dependence on the temperature, as illustrated
in Fig.~2 and further discussed below.
Using a field theoretical approach \cite{horovitz} it is argued
that the renormalized value  of the high frequency cutoff is
\beq{1052}
\omega_c \Big|_{\tbox{effective}} \ \ = \ \ \max\{\,\,\gamma,\,\,\Delta \,\,\}
\eeq
where $\gamma$ is the classical damping rate of Eq.(\ref{e1050}),
and $\Delta$ is the level spacing. The reasoning is as follows:
all the higher frequencies contribute to mass renormalization
only, and do not affect the dephasing process.
In more details: The significant renormalization starts
only below $\omega_c$ where the linear $|\omega|$ dispersion
of the dissipation term dominates and leads to $\ln \omega$ terms
in perturbation theory and to the need of either renormalization
group or an equivalent variational method \cite{horovitz}.

We would like to remark that if we do not
apply Eq.(\ref{e1052}), the results that we derive below are
affected quantitatively but not qualitatively.
Substitution of Eq.(\ref{e1052}) into Eq.(\ref{e1051}) implies
that ${\eta\ell^2 < 1}$ and ${\eta\ell^2 > 1}$
define distinct regimes of behavior.
For a dirty metal environment ${\eta\ell^2 \ll 1}$ is
equivalent to ${\alpha \ll 1}$ i.e. ${k_F\ell \gg 1}$.
For a CL environment $\eta\ell^2$ is formally infinite,
or one may say that $\ell$ is effectively determined
by the finite size $L$ of the system.

We come back to the dephasing rate calculation.
In the case of a fluctuating environment with
a short correlation distance~$\ell$,
we see that the high temperature (${T>T^*}$) result is:
\beq{28}
\hspace*{-15mm}
\Gamma_{\varphi} \ \ \approx \ \
\left\{\amatrix{
(2\eta \ell^2) \,  T
& \ \ \ \ \ \ \ \ \mbox{if $\eta\ell^2 \gg 1$}
\cr
(2\eta \ell^2)^{3/2} \, T
& \ \ \ \ \ \ \ \ \mbox{if $\eta\ell^2 \ll 1$}
}\right.
\ \ \ \ \ \ \ \ \Big[\mbox{for $T>T^{*}$}\Big]
\eeq
where in the ${\eta\ell^2\ll 1}$ expression we have
identified the effective (renormalized) cutoff as $\omega_c=\gamma$.
The strong coupling result (${\eta\ell^2 \gg 1}$)
cannot be trusted because the self-consistency
requirement (${\Gamma_{\varphi} \ll T}$) is not satisfied.
This is not in contradiction with the observation
that the CL result Eq.(\ref{e27}) is formally
a special case of the strong coupling result with ${\ell\mapsto L}$.
In the latter case the self consistency relation
becomes ${\eta L^2 \ll 1}$ irrespective of~$\ell$.

In the low temperature regime we have (be definition)
${q^{*}=q_T}$. Consequently the~$T$ dependence becomes non-linear,
and we get ${\Gamma_{\varphi} \approx \eta\ell^3M^{1/2} T^{3/2}}$.
The similarity of the latter to the familiar result for a
diffusive electron is misleading. In both cases $q_T\propto
T^{1/2}$ but for different reasons, and with a different
prefactors. For sake of completeness we write the precise
expression which is obtained for a dirty metal environment using
Eq.(\ref{e26}) with Eq.(\ref{e15}) and Eq.(\ref{e9}):
\beq{29} 
\hspace*{-15mm} \Gamma_{\varphi} \ \ = \ \
\frac{e^2}{4\pi^2\sigma} \, T \, q_T
\ln\left(\frac{1}{q_T\ell}\right) \ \ \ \ \ \ \ \ \ \ \ \ \ \ \ \
\ \ \Big[\mbox{for $\Delta < T < T^{*}$}\Big] 
\eeq
Since $\Gamma_{\varphi}\sim T^{3/2}$ at sufficiently low $T$
the condition $\Gamma_{\varphi}\ll T$ is valid even for strong
coupling $\eta \ell^2>1$. We note also that inclusion of a $q=0$
mode in (\ref{e26}) would have led to $\Gamma_{\varphi}\sim w_0T$.
Hence our precise formulation in Eq.(\ref{e1002}) is essential.

The crossover from the high temperature result to the low temperature
result is illustrated in Fig.~2.
The illustration assumes ${\eta \ell^2 \ll 1}$ which
implies that the self consistency requirement
(${\Gamma_{\varphi}\ll T}$) is globally satisfied.
It also should be realized that
the low temperature regime ${\Delta \ll T \ll T^{*}}$
exists only for a ``large ring" (${\eta L^2 \gg 1}$),
for which ${T^*=\gamma \gg \Delta}$.
Finally, for the $L<\ell$ case, the weight $w_{\pm1}$ dominates,
leading to $\mathcal{M}_{\tbox{eff}}=1$
and ${\Gamma_{\varphi}=2\eta L^2 T}$ as in Eq.(\ref{e27}),
which for $\omega_c<\Delta$ is consistent
with $\Gamma_{\varphi}<\Delta$ for $\eta L^2<1$.

%%%%%%%%%%%%%%%%%%%%%%%%%%%%%%%%%%%%%%%%%%%%%%%%%%%%%%%%%%%%%%%%
%%%%%%%%%%%%%%%%%%%%%%%%%%%%%%%%%%%%%%%%%%%%%%%%%%%%%%%%%%%%%%%%
\section{Dephasing in the presence of a dirty metal ``$T{=}0$"}

We would like to discuss the ``zero temperature" regime.
One should be very careful in specifying the conditions
of physical interest, else the problem may become trivial
or of no experimental relevance. In what follows we
assume that the dimensionless coupling between the
system and the environment ($\eta\ell^2$ for a dirty metal,
or $\eta L^2$ for a CL environment)
is much smaller than unity. This means that the competing
energy scales are the level spacing $\Delta \sim 1/(ML^2)$
and the temperature. So the simplest definition
of zero temperature is ${T \ll \Delta}$ for which
the system is in the ground state with an exponentially
small probability to be found in an excited state.
In this regime mass renormalization effect can be calculated
using second-order perturbation theory, as in the
Polaron problem, or possibly using field theoretical methods.
Furthermore in this regime we can treat the dephasing
problem using a ``two level approximation",
which is a very well studied model \cite{weiss}.

The notion of ``zero temperature" is also applicable
if ${T \gg \Delta}$ provided the time of interest
is short (${t \ll 1/T}$).
\hide{Note that ${1/T \sim m\mbox{Sec}}$ in the $\mu\mbox{K}$ regime.}
In this regime the power spectrum $\tilde{S}(q,\omega)$
is the same as for ${T{=}0}$ within the
frequency interval  ${T \ll \omega \ll \omega_c}$.
Consequently the Ohmic temporal correlations
are ${C(\tau) \approx -(\eta/\pi)/\tau^2}$
within the time interval ${(1/\omega_c) \ll t \ll (1/T)}$.
As explained in a previous section such correlations
may imply a logarithmic growth of $p_{\varphi}(t)$.
In view of the claim that the renormalized
value of $\omega_c$ is the damping rate~$\gamma$,
it follows that logarithmic spreading may arise
only if ${T \ll \gamma}$, which is the low temperature regime.

As discussed in a previous section, Eq.(\ref{e1001}) gives
a non-zero result for $p_{\varphi}(t)$ even at zero temperature.
For the CL model we have only $q{\sim}0$ fluctuating modes
and we get the expected $\log(t)$ spreading:
\beq{35}
p_{\varphi}(t) = \frac{\eta}{\pi} \left(\frac{L}{2\pi}\right)^2  \, \ln\left(\omega_c t\right)
\ \ \ \ \ \ \ \ \Big[\mbox{Caldeira-Leggett}\Big]
\eeq
where $\omega_c$ is the high frequency cutoff of the temporal fluctuations.
More generally, for a particle in a ring the result can be written
as a sum over all the $q$ Fourier components:
\beq{1002}
p_{\varphi}(t) \ \ = \ \ \frac{\eta}{\pi}
\sum_{m} w_m  \ln\left(\frac{\omega_c}{(1/t)+\omega(q_m)}\right)
\eeq
Strictly speaking for a finite system the assumption $T \gg \Delta$
always breaks down in the zero temperature limit.
Still it is meaningful to formulate a condition for {\em not} having dephasing
at zero temperature irrespective of the finite size effect:
\beq{0}
\lim_{L\rightarrow\infty} p_{\varphi}(t={\infty}) \ \ \ll \ \ 1
\eeq
Thus the question is simply whether in the continuum limit the $q$
summation in Eq.(\ref{e1002}) converges in its lower limit. For a
fluctuating environment with a finite (short) correlation
distance $\ell$,
\beq{36}
p_{\varphi} \ \ \sim \ \  \int_0^{1/\ell} dq \, \eta\ell^3 \, \ln\left[\frac{M\omega_c}{q^2}\right]
\ \ \sim \ \  \eta \ell^2 \ \ln\left(\frac{\omega_c\ell^2}{M}\right)
\eeq
We conclude that for a dirty metal environment with ${\ell\ll L}$,
coherence is maintained if $\eta \ell^2\ll 1$, i.e. if $k_F\ell \gg 1$.

It is important to notice the following: in the strict Caldeira-Leggett
limit (${\ell=\infty}$) the size of the ring~$L$ can be arbitrarily large,
hence the Heisenberg time $1/\Delta$ becomes huge, and Eq.(\ref{e1002})
leads to Eq.(\ref{e35}), which becomes an exact result.
So if we consider a {\em dirty metal environment}
with long wavelength fluctuations (${\ell \gg L}$),
it looks as if we are back in the ``CL regime"
leading to Eq.(\ref{e35}).
But this is not quite correct unless we give away the weak
coupling assumption ${\eta\ell^2 \ll 1}$. As long as we keep
$\alpha \ll 1$ (fixed) the CL result does not apply.
This is because once $\ell\rightarrow L$ and ${\eta L^2 < 1}$
the quantization of the energy spectrum becomes important and
the renormalized cutoff frequency Eq.(\ref{e1052}) becomes
${\omega_c \sim \Delta}$ instead of ${\omega_c \sim \gamma}$.
Accordingly, in the latter case, the time during which the log
spreading prevails diminishes.

%%%%%%%%%%%%%%%%%%%%%%%%%%%%%%%%%%%%%%%%%%%%%%%%%%%%%%%%%%%%%%%%
%%%%%%%%%%%%%%%%%%%%%%%%%%%%%%%%%%%%%%%%%%%%%%%%%%%%%%%%%%%%%%%%
\section{Mass renormalization}

It can be shown \cite{dld} that a particle that interacts
with a fluctuating `dirty' environment acquires an additional
inertial (polaronic) mass.
However in recent works \cite{guinea,golubev,horovitz}
the  mass renormalization concept appears in a new context.
The free energy $\mathsf{F}(T,\Phi)$
of a particle in a ring is calculated, where $T$ is
the equilibrium temperature and $\Phi$ is the Aharonov Bohm flux
the through the ring. Then the coherence is characterized by
the ``curvature", which is a measure for the sensitivity to $\Phi$.
If the interaction with the environment is
negligible the result can be written as
\beq{37}
\left.\frac{\partial^2 \mathsf{F}}{\partial\Phi^2}\right|_{\Phi{=}0}
\ \ = \ \ \frac{e^2}{M^*L^2}f(M^{*}L^2T)
\eeq
with the bare mass $M^*=M$.
The dependence of the curvature on~$T$ merely
reflects the Boltzmann distribution of the energy.
In the presence of coupling to the environment
it turns out that $M^*>M$.
At $T=0$, for fixed $\eta\ell^2 \ll 1$, Monte Carlo data show
\cite{horovitz2} that the ratio $M^*/M$ is independent of the
radius beyond a critical $L_c$.
As the radius becomes smaller than $L_c$,
the ratio $M^*/M$ rapidly approaches unity.
In the regime of ``large $L$" the mass renormalization
effect diminishes with the temperature
and depends on the scaled variable $LT$,
while for ``small $L$" the ratio $M^*/M$
grows with the temperature, and depends on
the scaled variable $L^4T$.

The natural question is whether we can shed some light
on the physics behind this observed temperature dependence
of the mass renormalization factor.
In particular we would like to explain why in in one
regime $M^*/M$ is a function of $LT$, while
in another regime it is a function of $L^4T$.
Making the conjecture that the temperature dependence
of $M^*/M$ is determined by dephasing it is natural
to suggest the following measure of coherence:
\beq{38}
x(T,L) \ \ = \ \
p_{\varphi}\left(t{=}\frac{1}{\Delta_{\tbox{eff}}}\right)
\ \ = \ \ \frac{\Gamma_{\varphi}}{\Delta_{\tbox{eff}}}
\eeq
Namely, it is the dephasing factor at the
time $t=1/\Delta_{\tbox{eff}}$, where $\Delta_{\tbox{eff}}$ is the
``relevant" energy scale. Equivalently the condition $x \ll 1$
means that the energy levels near $\Delta_{\tbox{eff}}$ remain sharp.
The inequality $x<1/2$ can serve as a practical definition
for having coherence. It can be translated either
as a condition on the temperature, or optionally it can be used
in order to define a coherence length that depends on the temperature.
The conjecture is that $y=M^*/M$ is a function of $x$.
Let us calculate $x$ using Eq.(\ref{e26}).
We assume $\eta\ell^2\ll 1$ but $\eta L^2$ can be either larger or smaller
compared to unity. This is equivalent to saying that the damping
rate~$\gamma$ can be either larger or smaller compared with the
level spacing $\Delta$. Using Eq.(\ref{e1051}) with
Eq.(\ref{e1052}) this further implies
that $\mathcal{M}_{\tbox{eff}}$ is either larger or of order unity
respectively. The transitions that are associated with
the ``relevant" energy levels are characterized
by ${q \sim q_{\tbox{eff}}}$ and
accordingly ${\Delta_{\tbox{eff}} \sim \mathcal{M}_{\tbox{eff}} \times (ML^2)^{-1}}$.
Using Eq.(\ref{e26}) we deduce that the result does not depend
on $\mathcal{M}_{\tbox{eff}}$ but only on $\bar{w}$, leading to
\beq{0}
x(T,L) \ \ = \ \ \eta \bar{w} M L^2 T
\ \ = \ \
\left\{\amatrix{
\eta M \ell^3 L T
& \ \ \ \ \ \ \ \ \ \mbox{for $L \gg \ell$}
\cr
\eta M L^4 T
& \ \ \ \ \ \ \ \ \ \mbox{for $L \ll \ell$}
}\right.
\eeq
We recall that our $\Gamma_{\varphi}$ is valid at least for
weak coupling $\eta\ell^2\ll 1$. The scaling of $\Gamma_{\varphi}$
with $L$ is consistent with Monte Carlo exponents for the
coherence length $L\sim T^{-\mu}$ with either $\mu=1$ or $\mu=1/4$.
The Monte Carlo data has not determined so far whether the
transition between the two regimes is at $L\approx \ell$
or whether it is coupling dependent.

%%%%%%%%%%%%%%%%%%%%%%%%%%%%%%%%%%%%%%%%%%%%%%%%%%%%%%%%%%%%%%%%
\section{Summary and Discussion}

In this paper we have defined and calculated the dephasing factor
$P_{\varphi}(t)$ for a particle in a ring
due to the fluctuations of a dirty metal environment.
At finite temperature we have calculated the dephasing rate
$\Gamma_{\varphi}$. Our interest was mainly in the mesoscopic
regime ${\Delta\ll\Gamma_{\varphi}\ll\gamma}$, where interference
is important (because ${\Gamma_{\varphi}\ll\gamma}$). Unlike the
microscopic regime (${\Gamma_{\varphi}\ll\Delta}$), which is
customary in atomic physics studies, here the question of
dephasing at low temperature is tricky both conceptually and
technically.

The decoherence is induced because the system
gets entangled with the environmental modes. It should be
clear that in generic circumstances the coupling always induces
``transitions" that lead to system-bath entanglement. Accordingly
we have ${P_{\varphi}(t)<1}$ even if ``${T=0}$". This by itself
does not mean ``having dephasing": entanglement is also associated
with the adiabatic renormalization due to the interaction with the
high frequency modes. In order to ``have dephasing" the loss of
purity should not be just a {\em transient}: rather it should be a
progressive process.

Still even with this careful point of view, the reader may doubt
whether the notion of ``dephasing factor" is really helpful in
studying dephasing. After all what do we ``really" mean by
dephasing. Possibly the ``correct" procedure is to study an
equilibrium correlation function $C(t)$, and to ask whether it is
damped in the $t\rightarrow\infty$ limit. In the absence of
coupling to the environment the Fourier transform
$\tilde{C}(\omega)$ is a sum over delta functions
${\delta(\omega-\Omega_r)}$. Due to the coupling the deltas are
broadened into resonances of with $\Gamma_r$. This is true at any
temperature, also at ``${T=0}$". The controversy about dephasing
at ``${T=0}$" is related to the limit $L\rightarrow\infty$. Do the
resonances overlap in this limit? Do singular features of the
uncoupled system survive? For sub-Ohmic bath \cite{weiss} 
the ratio $\Gamma_{\varphi}/\Delta$, where $\Delta$ is the mean
level spacing, diverges as $L\rightarrow\infty$. But the Ohmic
case is ``marginal" and within the framework of the
Fermi-golden-rule it remains a constant $\alpha$. So if this
$\alpha$ is smaller compared with unity, we naively expect no
dephasing at ``${T=0}$".

The naive expectation of having no dephasing at ``${T=0}$"
is not without loopholes. One obvious loophole is the mass
renormalization issue. If hypothetically the renormalized
mass and hence the density-of-states diverge as ${T\rightarrow0}$,
it might imply dephasing at zero temperature. The recent studies of
equilibrium properties of the ring problem are aimed in
studying this question carefully, in a controlled way.
For a particle that interacts with a dirty metal environment
we believe, on the basis of \cite{horovitz,horovitz2} that
the renormalized mass at zero temperature is finite.
So in the case of a dirty metal environment there is no indication
for ``dephasing at $T{=}0$".

Still one would like to know what happens at low but finite
temperature. As we said previously, no doubt that study of
equilibrium properties is conceptually the
best procedure. Still, we also want to physically
understand the results. Here we come back to the `dephasing
factor' notion. In spite of the problems which are associated with
this concept, we believe that it is powerful enough to shed light
on the physics of dephasing. 
Our aim in this paper was to maximally exploit this notion, within the
Fermi golden rule picture, in order to demonstrate that it
captures the correct physics of all the established results
regarding dephasing. In particular it has provided an an
explanation for the $T$~dependence of the mass renormalization 
effect, and under what conditions the spatial aspect of the
fluctuations is capable of suppressing the ``$T{=}0$" power law
decay of coherence.

%%%%%%%%%%%%%%%%%%%%%%%%%%%%%%%%%%%%%%%%%%%%%%%%%%%%%%%%%%%%%%%%%%%%%%%%%%%%%%%%%%%%%%%%%%%
%%%%%%%%%%%%%%%%%%%%%%%%%%%%%%%%%%%%%%%%%%%%%%%%%%%%%%%%%%%%%%%%%%%%%%%%%%%%%%%%%%%%%%%%%%%

\appendix

%%%%%%%%%%%%%%%%%%%%%%%%%%%%%%%%%%%%%%%%%%%%%%%%%%%%%%%%%%%%%%%%
\section{Modeling of a fluctuating environment}

It is customary to write the system-environment Hamiltonian as
\beq{0}
\mathcal{H}_{\tbox{total}} \ \ = \ \
\mathcal{H}_{\tbox{sys}} (\hat{x},\hat{p}) +
V_{\tbox{int}}(\hat{x},\hat{Q}_{\alpha}) +
\mathcal{H}_{\tbox{env}}(\hat{Q}_{\alpha},\hat{P}_{\alpha})
\eeq
where $(\hat{x},\hat{p})$ are the canonical coordinates
of the particle, and $(\hat{Q}_{\alpha},\hat{P}_{\alpha})$
are the environmental degrees of freedom.
In the case of an interaction of a particle
with a dirty metal environment (in 3~dimensions):
\beq{0}
V_{\tbox{int}}
\ \ = \ \
\int d^3x \hat{\rho}(\bm{x})
\int d^3x'
\frac{e^2 \hat{\mathsf{n}}(\bm{x}') }{|\bm{x}-\bm{x}'|}
\ \ \equiv \ \
\int d^3x' \hat{\rho}(\bm{x}) \, \hat{\mathcal{U}}(\bm{x})
\eeq
where the electronic density $\mathsf{n}(\bm{x})$
can be expressed as a function of their coordinates,
while ${\hat{\rho}(\bm{x}) =  \delta(\bm{x}-\hat{\bm{x}})}$
is a particle related field operator.

In order to allow a Feynman-Vernon treatment it is more
convenient to regard $\mathcal{U}(\bm{x})$ as arising from
the interaction with a bath of harmonic
oscillators \cite{dld}. Each harmonic oscillator is a scatterer
which is characterized by its location $x_{\alpha}$ and its
natural frequency $\omega_{\alpha}$. The interaction of the
particle with the $\alpha$ scatterer is ${\hat{Q}_{\alpha} \,
u(\hat{x}-x_{\alpha})}$, so we write (in one dimension):
\beq{0}
V_{\tbox{int}}  \ \ = \ \
\sum_{\alpha} c_{\alpha} \hat{Q}_{\alpha} \, u(\hat{x}-x_{\alpha})
\ \ = \ \
\int dx \, \hat{\rho}(x) \, \hat{\mathcal{U}}(x)
\eeq
where the $c_{\alpha}$ are coupling constants,
and $\hat{\rho}(x)=\delta(\hat{x}-x)$.
In the Heisenberg (interaction) picture
a time index is added so we have $\hat{\rho}(x,t)$ and $\hat{\mathcal{U}}(x,t)$.
Accordingly the fluctuating filed is

\beq{2}
\hat{\mathcal{U}}(x,t) \ \ = \ \ \sum_{\alpha} c_{\alpha} \hat{Q}_{\alpha}(t) \, u(x-x_{\alpha})
\eeq
As explained in Ref.\cite{dld} it is possible to postulate the
interaction $u(r)$, and the distribution of the parameters
$(x_{\alpha},\omega_{\alpha},c_{\alpha})$, such as to obtain a
fluctuating field with a physically desired $\tilde{S}(q,\omega)$.
This type of modeling is equivalent to the field-theoretical
assumption of having Gaussian fluctuations, and accordingly
a linear response treatment of the environment becomes exact.

%%%%%%%%%%%%%%%%%%%%%%%%%%%%%%%%%%%%%%%%%%%%%%%%%%%%%%%%%%%%%%%%
\section{The fluctuations of a dirty metal}

For a metal we can use FD realation in order to relate
the spati-temporal power spectrum $\tilde{S}(q,\omega)$ to the conductivity:
\beq{100}
\tilde{S}(\bm{q},\omega) =
\frac{4\pi e^2}{\bm{q}^2} \im\left[ \frac{-1}{\varepsilon(\bm{q},\omega)} \right]
\frac{2\hbar}{1-\eexp{-\hbar\omega/T}}
\eeq
where
\beq{101}
\varepsilon(\bm{q},\omega) = 1 + \frac{4\pi\sigma}{-i\omega+D\bm{q}^2}
\eeq
and
\beq{102}
\im\left[ \frac{1}{\varepsilon(\bm{q},\omega)} \right] =
-\frac{4\pi\sigma\omega}{(D\bm{q}^2+4\pi\sigma)^2 + \omega^2}
\eeq
Thus we get
\beq{103}
\tilde{S}(\bm{q},\omega) \approx
\frac{e^2}{\sigma}
\frac{1}{\bm{q}^2}
\frac{2\hbar\omega}{1-\eexp{-\hbar\omega/T}}
\ \ \ \ \ \ \mbox{for}
\ \ |\omega| \lesssim \omega_c,
\ \  |\bm{q}|\lesssim \frac{1}{\ell}
\eeq
The ohmic behavior is cut-off by the
Drude collision frequency $\omega_c$,
and the elastic mean free path is $\ell=v_F/\omega_c$,
where $v_F$ is the Fermi velocity.
The expression for $\eta$ in Eq.(\ref{e9}) is
obtained from the Kubo formula Eq.(\ref{e6}).

%%%%%%%%%%%%%%%%%%%%%%%%%%%%%%%%%%%%%%%%%%%%%%%%%%%%%%%%%%%%%%%%
\section{Calculation of Fourier components}

We are interested only in fluctuations within the ring.
Therefore we have to calculate the Fourier components
of the correlator ${\big\langle \hat{\mathcal{U}}(x(\theta_2),t_2) \hat{\mathcal{U}}(x(\theta_1),t_1) \big\rangle}$. For the CL model we use the integral
\beq{201}
\hspace*{-15mm}
\int d^3\bm{q} \, \frac{3\delta^3(\bm{q})}{\bm{q}^2}  \, \eexp{i\bm{q}\cdot \bm{R}}
 =  \const - \frac{1}{2}\bm{R}^2
 =  \const + \left(\frac{L}{2\pi}\right)^2 \cos(\theta_2{-}\theta_1)
\eeq
where $\bm{R}=\bm{R}(\theta_2){-}\bm{R}(\theta_1)$ 
so that $|\bm{R}|=|2\sin((\theta_2{-}\theta_1)/2)|[L/2\pi]$.
For a dirty metal with fluctuations within $q\lesssim 1/\ell$ we have
\beq{202}
\hspace*{-15mm}
\int \frac{d^3\bm{q}}{(2\pi)^3} \,\frac{4\pi\ell^3}{\bm{q}^2} \,
\eexp{i\bm{q}\cdot \bm{R}}
= \frac{\ell^3}{\sqrt{\bm{R}^2 + \ell^2}}
= \ell^2\left[a_0+\sum_{m=1}^{\infty} a_m \cos(m(\theta_2{-}\theta_1)) \right]
\eeq
Our $a_{m{\ne}0}$ are half the ``convention" in Ref.\cite{guinea}.
From the Fourier transform relation
it follows that $\sum_{m=0}^{\infty} a_m =1$, and we also have the
sum rule $\sum_{m=1}^{\infty} a_m m^2 = \mathcal{M}^2$
where $\mathcal{M}=(L/(2\pi))/\ell$.
Disregarding the $m=0$ Fourier component the following approximation
can be obtained \cite{guinea,golubev} for ${\mathcal{M}\gg 1}$
\beq{203}
a_m \approx \frac{1}{\pi\mathcal{M}}
\,\ln\left(\frac{\mathcal{M}}{m}\right) \ \ \ \ \ \ \ \ \ \
\mbox{for $0 < m < \mathcal{M} $}
\eeq
From Eq.(\ref{e10001}) it follows that ${w_m=\ell^2a_m/2}$.
We conclude that the particle in the ring experiences {\em white} fluctuating field
that is characterized by a correlation distance $\ell$. The fluctuating field can
be re-interpreted as arising from a short range interaction $u(r)$ with uniformly
distributed set of scatterers as in Eq.(\ref{e2}).

In the other extreme case of CL-like environment (${\ell \gg L}$)
the fluctuations of the higher (${m>1}$) modes
are negligible compared with the
fluctuations of the lowest (${m{=}1}$) mode.
Accordingly we say that the number of effective modes is ${\mathcal{M}=1}$.
Using the ad-hock notation ${\bar{\mathcal{M}}=(L/(2\pi))/\ell \ll 1}$,
the sum rule which is based on Eq.(\ref{e202})
implies that $a_1=\bar{\mathcal{M}}^2$ while $a_{m>1}$
have higher powers of~$\bar{\mathcal{M}}$.

%%%%%%%%%%%%%%%%%%%%%%%%%%%%%%%%%%%%%%%%%%%%%%%%%%%%%%%%%%%%%%%%
\section{The dephasing factor - semiclassical perspective}

For short range scattering, if $x^A$ and $x^B$ of Eq.(\ref{e16})
are well separated, and hence interact with {\em different}
sets of oscillators, we can argue within the semiclassical
framework that $P_{\varphi}$ is the probability to induce
an excitation in the bath (i.e. ``to leave a trace in the environment").
The argument is elaborated in Appendix~C of \cite{qbm}.
This argument {\em fails} if the interfering states are not
well separated in space, but rather interact with the {\em same}
oscillators. For this reason the definition of the dephasing factor
has to be refined. One possibility is to adopt a ``scattering"
point of view, hence treating correctly closed channels.
Using sloppy notations the idea is to define the dephasing factor
in analogy with Eq.(\ref{e16}) as
${P_{\varphi} = \Big|\Big\langle U[\psi^A] \chi \Big| U[\psi^B] \chi \Big\rangle\Big|}$
where $\psi^A$ and $\psi^B$ are {\em ingoing states} of the system.
This way of writing is suggestive rather than exact.
Referring to a superposition preparation of the ring,
where $\psi^A$ and $\psi^B$ are momentum eigenstates,
it is clear that $P_{\varphi}$ is not necessarily
the same as the probability to induce an excitation in the bath.
This is because $U[\psi^A]$ and $U[\psi^B]$ involve
the excitation of the {\em same} oscillators, rather
than {\em different} sets of oscillators.
If the factorized preparation is ${|p_0 n_0\rangle}$,
then we write the evolved state in the interaction picture
after time $t$ as ${|(p_0 n_0)_t\rangle}$.
If we have initially a superposition ${|p_1\rangle+|p_2\rangle}$,
the evolved state would be
\beq{301}
|\Psi\rangle \ \ = \ \ |p_1\rangle \otimes |\chi^{(1)}\rangle \ + \ |p_2\rangle \otimes |\chi^{(2)}\rangle \ + \ \mbox{InelasticPart}
\eeq
where the so-called relative states of the bath are
\beq{302}
\chi^{(1)}_n \ \ = \ \ \langle p_1 n|(p_1 n_0)_t\rangle \\
\chi^{(2)}_n \ \ = \ \ \langle p_2 n|(p_2 n_0)_t\rangle
\eeq
The dephasing factor is
\beq{303}
P_{\varphi}  =  |\langle \chi^{(1)}|\chi^{(2)}\rangle|
 =   P_t(p_0,n_0|p_0,n_0) + \sum_{n(\ne n_0)} {\chi^{(1)}_n}^{*} \chi^{(2)}_n
\eeq
where we assume ${p_1 \sim p_2 \sim p_0}$.
It is not difficult to realize that the same approximation
implies that the second term equals  $P_t(p_0,n{\ne}n_0|p_0,n_0)$
in agreement with Eq.(\ref{e18}). With some further argumentation
we can justify the the second term in Eq.(\ref{e18}) as well.
We note that this derivation parallels the semiclassical treatment
in Appendix~D of~\cite{qbm}, where $\tilde{P}(q,\omega)$ is defined as the
difference ${\tilde{P}_{\parallel}(q,\omega)-\tilde{P}_{\perp}(q,\omega)}$.

%%%%%%%%%%%%%%%%%%%%%%%%%%%%%%%%%%%%%%%%%%%%%%%%%%%%%%%%%%%%%%%%
\section{The purity based definition of the dephasing factor}

In this appendix we explain the derivation of Eq.(\ref{e18}) from
Eq.(\ref{e17}). The zero order term in Eq.(\ref{e17}) is
the ${p'{=}p''{=}p_0, n'{=}n''{=}n_0}$ term. It is equal to $P_0^2$
where $P_0=P_t(p_0,n_0|p_0,n_0)$. There are four sets of first
order terms:
The sum of the ${p'{=}p''{=}p_0, n'{=}n_0,n''{\ne}n_0}$ terms
is ${P \times p_{\tbox{sys}}}$
where ${p_{\tbox{sys}} = P_t(p_0,n{\ne}n_0|p_0,n_0) }$.
Here $n\neq n_0$ implies a summation $\sum_{n\neq n_0}$.
The sum of the ${p'{=}p''{=}p_0, n'{\ne}n_0, n''{=}n_0}$ terms
is the same. There are two other sets,
with either $p'\neq p_0$ or $p''\neq p_0$,
that give each ${P \times p_{\tbox{env}}}$,
where ${p_{\tbox{env}} = P_t(p{\ne}p_0,n_0|p_0,n_0)}$.
Summing over all the leading order contributions we get
\beq{0}
P_{\varphi} \ \ = \ \ \Big[ P_0^2 + 2P_0 \times p_{\tbox{sys}} +  2P_0 \times p_{\tbox{env}}  + \mathcal{O}(p^2) \Big]^{1/2}
\eeq
leading to $P_{\varphi}\approx P_0 + p_{\tbox{sys}} + p_{\tbox{env}}$
which is Eq.(\ref{e18}).

%%%%%%%%%%%%%%%%%%%%%%%%%%%%%%%%%%%%%%%%%%%%%%%%%%%%%%%%%%%%%%%%%%%%%%%%%%%%%%%%%%%%%%%%%%%
%%%%%%%%%%%%%%%%%%%%%%%%%%%%%%%%%%%%%%%%%%%%%%%%%%%%%%%%%%%%%%%%%%%%%%%%%%%%%%%%%%%%%%%%%%%

\clearpage

\ \\ \ \\

{\bf Acknowledgment:} \
We thank Florian Marquardt, Jan von Delft, and Joe Imry 
for discussions and helpful communications.
This research was supported by a grant from the DIP,
the Deutsch-Israelische Projektkooperation.

\ \\ \ \\

%%%%%%%%%%%%%%%%%%%%%%%%%%%%%%%%%%%%%%%%%%%%%%%%%%%%%%%%%%%%%%%%%%%%%%%%%%%%%%%%%%%%%%%%%%%
%%%%%%%%%%%%%%%%%%%%%%%%%%%%%%%%%%%%%%%%%%%%%%%%%%%%%%%%%%%%%%%%%%%%%%%%%%%%%%%%%%%%%%%%%%%

%%%%%%%%%%%%%%%%%%%%%%%%%%%%%%%%%%%%%%%%%%%%%%%%%%%%%%%%%%%%%%%%
%%%%%%%%%%%%%%%%%%%%%%%%%%%%%%%%%%%%%%%%%%%%%%%%%%%%%%%%%%%%%%%%
% FIGURES

\clearpage

\ \\ \ \\

\putgraph[width=0.9\hsize]{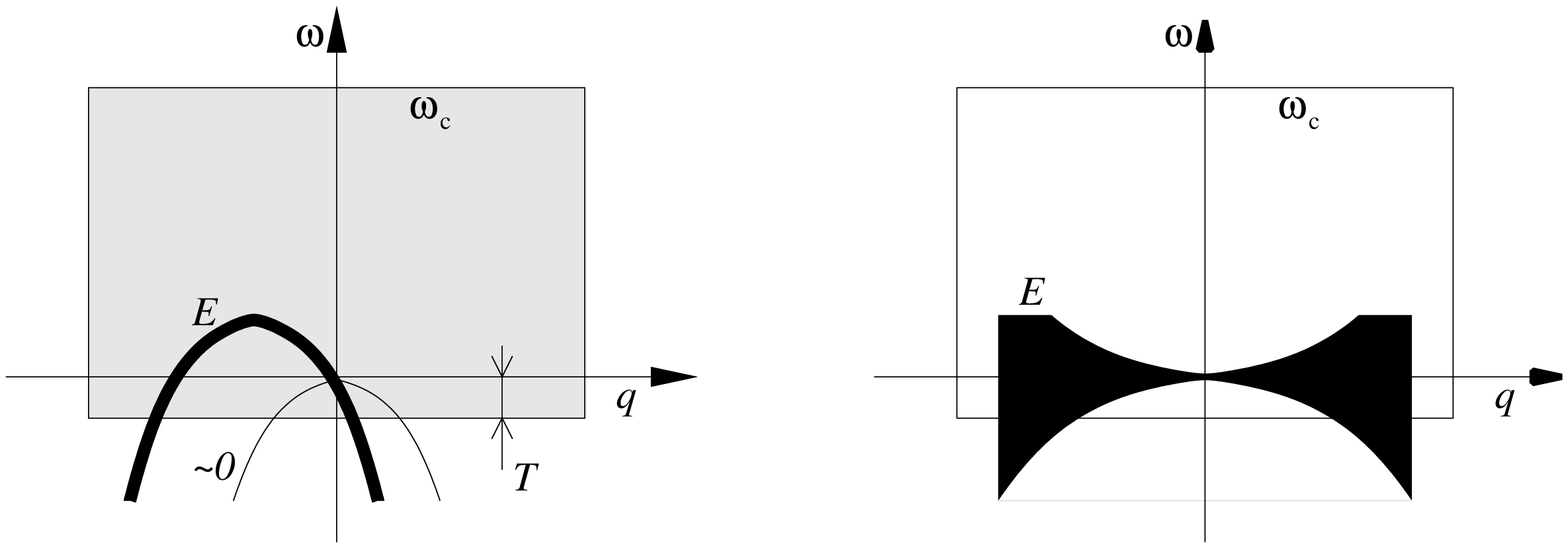}

{\footnotesize {\bf Fig.1:} The $(q,\omega)$ plane. The power spectrum $\tilde{S}(q,\omega)$
is distributed pre-dominantly within the rectangular area $q\lesssim(1/\ell)$,
that has a high frequency absorption cutoff~$\omega_c$, and a lower emission cutoff.
The emission cutoff~$T$ in this illustration reflects an assumption of having ${T<\omega_c}$,
otherwise it would be equal $\omega_c$ too.
The power spectrum $\tilde{P}(q,\omega)$ which is associated with the ballistic motion (left panel)
or with the diffusive motion (right panel) of the particle is illustrated by the
dark region. In both cases the energy~$E$ of the particle implies a frequency cutoff,
which is analogous to $T$. Close to equilibrium one should take ${E \sim T}$,
but without much error we take for low temperatures ballistic motion ${E\sim 0}$,
which is also illustrated in the left panel. }

\ \\ \ \\

\putgraph[width=0.6\hsize]{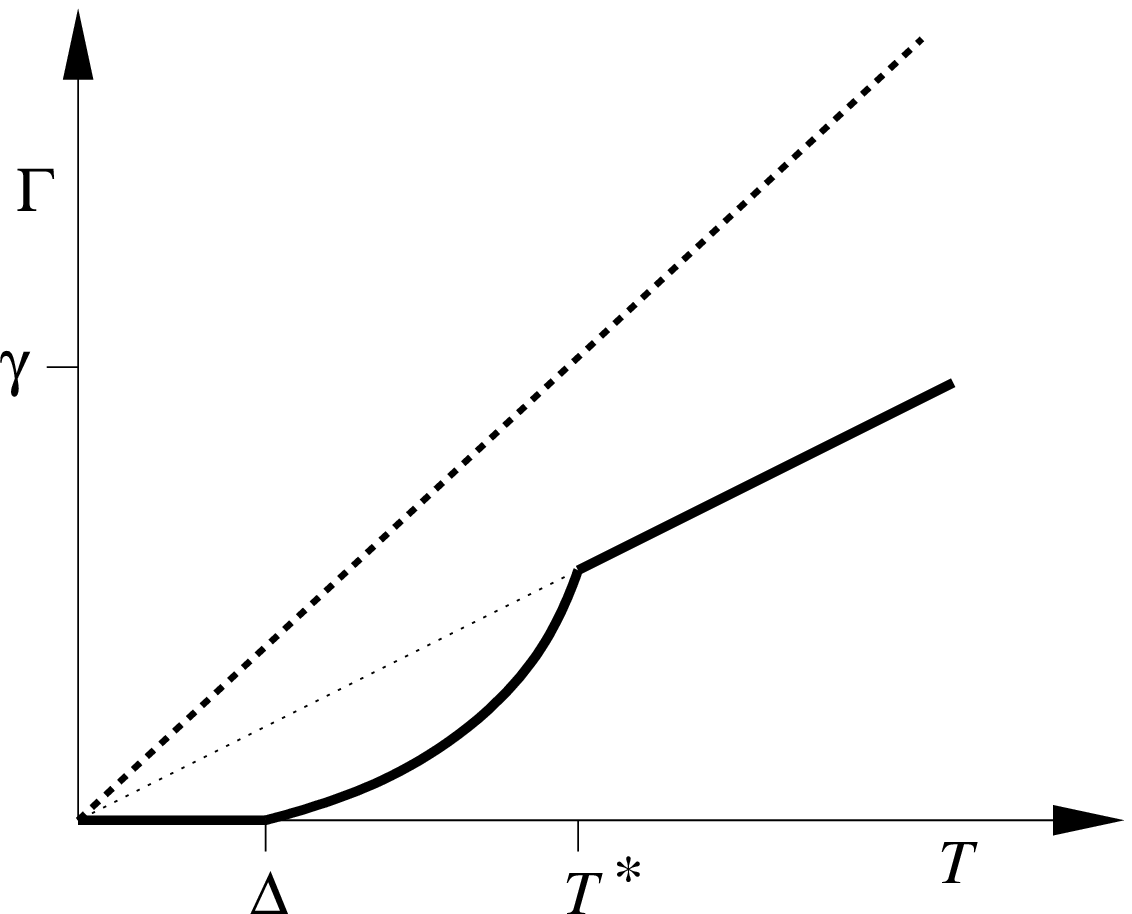}

{\footnotesize {\bf Fig.2:} Illustration of the dependence of the dephasing
rate~$\Gamma$ on the temperature~$T$.  The dephasing rate is well defined
for ${t>(1/T)}$, and hence the self consistency requirement is $\Gamma\ll T$.
This condition is demonstrated by a comparison with the dashed line.
The illustration reflects an assumption of having ${\eta\ell^2 \ll 1}$,
and therefore the crossover temperature~$T^*$ is equal to the damping
rate ${\gamma=\eta/M}$. The illustration further reflects an assumption
of ``large ring" (${\eta L^2 \gg 1}$) for which ${\gamma\gg\Delta}$,
else the low temperature regime (${\Delta \ll T \ll T^{*}}$) disappears.
For extremely low temperatures, such that $T$ is smaller compared
with the spacing ${\Delta = 1/(ML^2)}$, the probability to excite the system
is exponentially small and the familiar two-level modeling becomes applicable.}

%%%%%%%%%%%%%%%%%%%%%%%%%%%%%%%%%%%%%%%%%%%%%%%%%%%%%%%%%%%%%%%%
%%%%%%%%%%%%%%%%%%%%%%%%%%%%%%%%%%%%%%%%%%%%%%%%%%%%%%%%%%%%%%%%
\end{document}